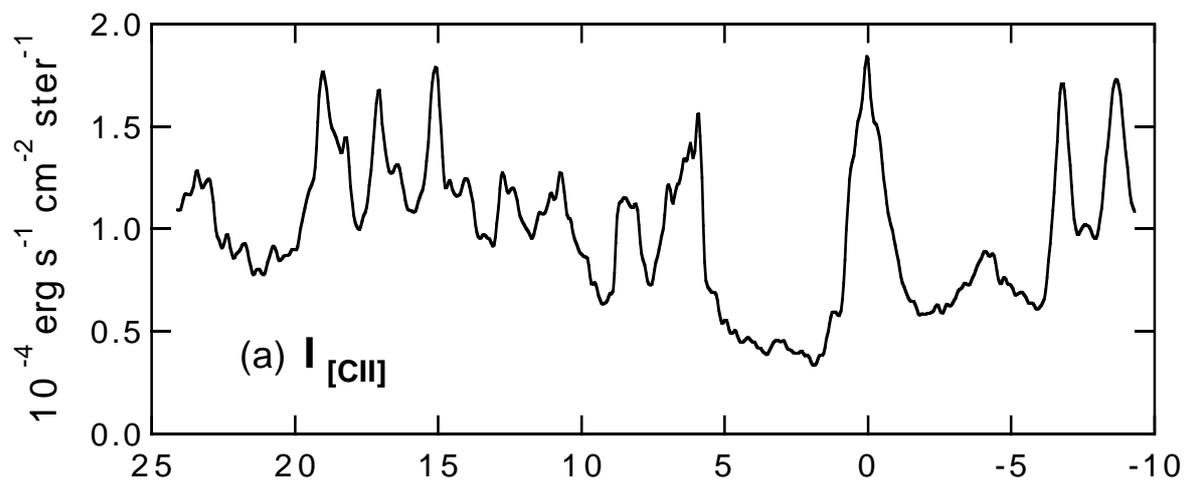
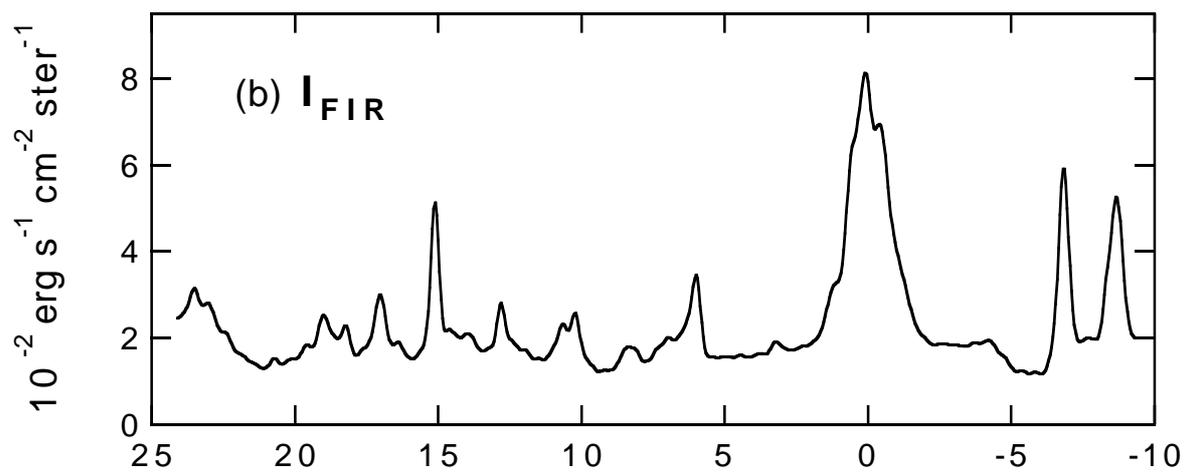
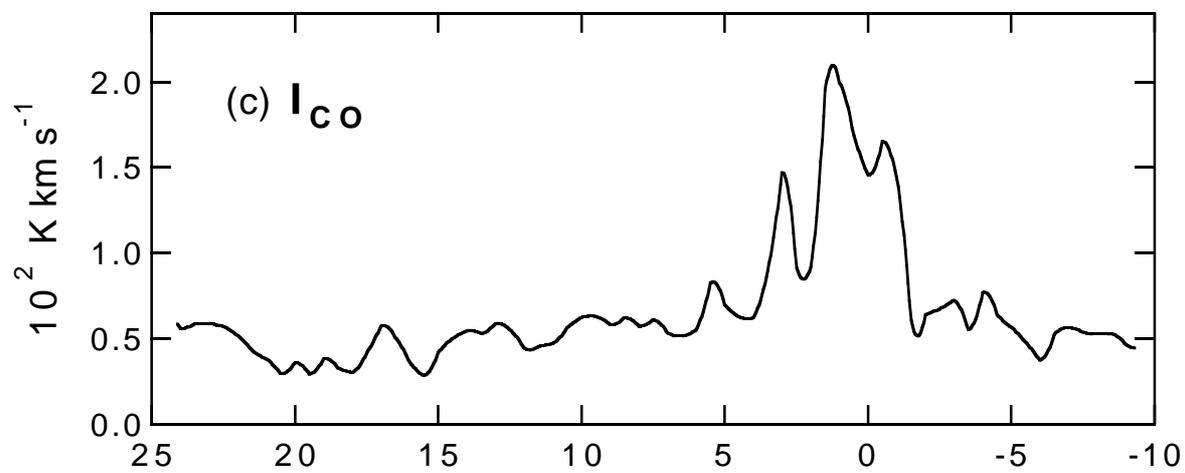
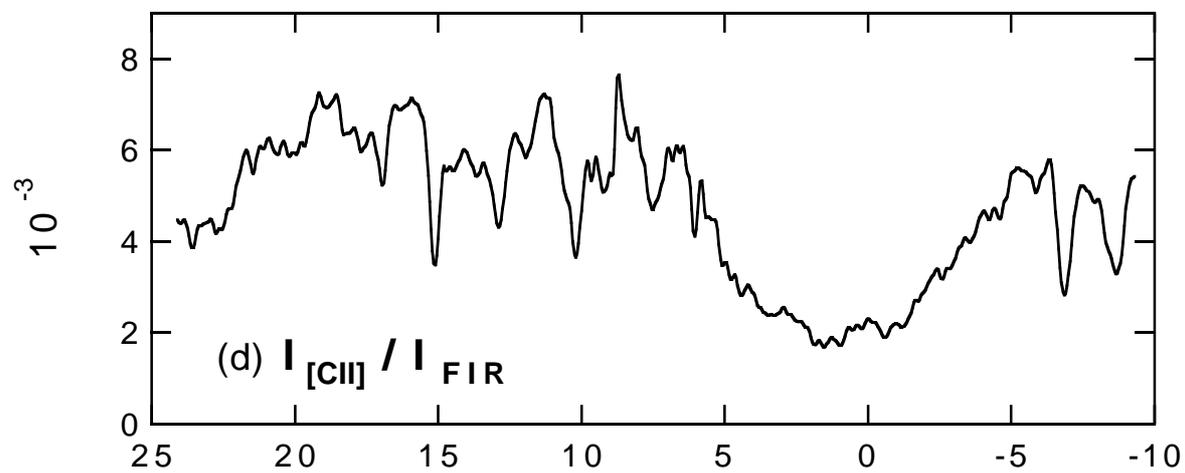

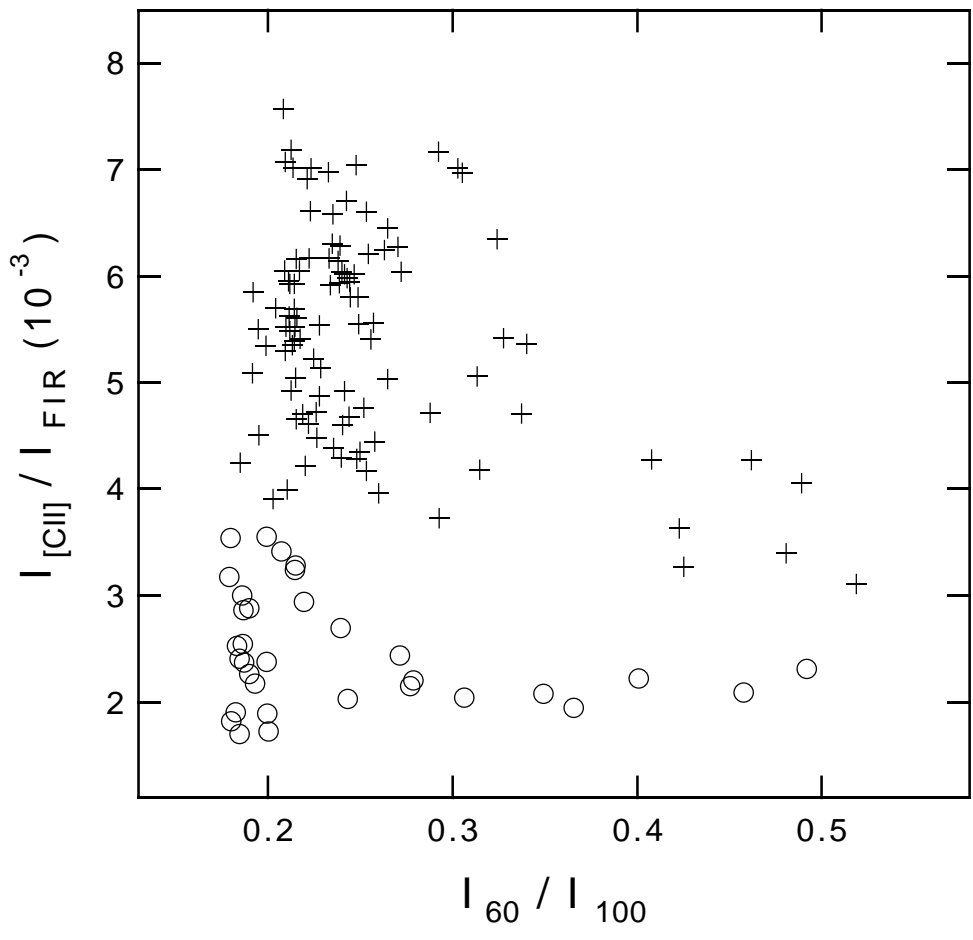

# Deficit of Far-Infrared [C II] Line Emission toward the Galactic Center

Takao Nakagawa, Yasuo Doi[1], Yukari Yamashita Yui[1,2], Haruyuki Okuda,

Kenji Mochizuki[1], and Hiroshi Shibai

The Institute of Space and Astronautical Science,

Yoshinodai 3-1-1, Sagamihara, Kanagawa 229, Japan

and

Tetsuo Nishimura[3] and Frank J. Low

Steward Observatory, University of Arizona, Tucson, AZ 85721

[1]Department of Astronomy, The University of Tokyo, Hongo 7-1-1, Bunkyo-ku, Tokyo 113, Japan

[2]Present Address:  Communications Research Laboratory, Nukui-kitamachi 4-2-1, Koganei, Tokyo 184, Japan

[3]Present Address: National Astronomical Observatory, Osawa 2-21-1, Mitaka, Tokyo 181, Japan

## ABSTRACT

We have observed the [C II] 158 $\mu$m line emission from the Galactic plane (-10°< $l$ < 25°, $|b| \leq 3°$) with the Balloon-borne Infrared Carbon Explorer (BICE). The observed longitudinal distribution of the [C II] line emission is clearly different from that of the far-infrared continuum emission; the Galactic center is not the dominant peak in the [C II] emission. Indeed, the ratio of the [C II] line emission to far-infrared continuum ($I_{\rm [CII]}/I_{\rm FIR}$) is systematically low within the central several hundred parsecs of the Galaxy.

The observational results indicate that the abundance of the $C^+$ ions themselves is low in the Galactic center. We attribute this low abundance mainly to soft UV radiation with fewer C-ionizing photons. This soft radiation field, together with the pervasively high molecular gas density, makes the molecular self-shielding more effective in the Galactic center. The self-shielding further reduces the abundance of $C^+$ ions, and raises the temperature of molecular gas at the $C^+/C/CO$ transition zone.

*Subject headings:* dust, extinction – Galaxy: center – infrared: ISM: lines and bands – ISM: molecules

## 1. Introduction

There is a substantial concentration of interstellar material within the central 500 pc of the Galaxy. (The distance to the Galactic center is assumed to be 8.5 kpc throughout this letter). The mass of molecular material in this region is $10^8 M_\odot$, which corresponds to about 10 % gas of the total molecular material in the whole Galaxy (Güsten 1989). The characteristics of interstellar gas in the center is quite different from those of interstellar gas in the Galactic disk (Güsten 1989): (1) the interstellar gas at the center is dominantly molecular, and (2) the clouds there are denser and warmer ($n \sim 10^4$ cm$^{-3}$ and $T \sim 70$ K) than typical clouds in the disk ($n \sim 10^{2.5}$ cm$^{-3}$ and $T \sim 15$ K).

Some of these characteristics can be naturally explained. The clouds must be sufficiently dense to withstand the tidal stress caused by the steep potential well in the region (Stark & Bania 1986), and the pervasively high density makes molecular forms more preferable than atomic forms.

However, the origin of the high temperature and the source of the luminosity, especially on a large scale, is still controversial (e.g., Morris 1989). Although there is clear evidence for current star-formation activity near the cores of Galactic center clouds (e.g., Sgr B$_2$), the relative paucity of the usual indicators of star-formation activity (e.g., H$_2$O masers, and SNRs) suggests that the star-formation rate per unit mass in the Galactic center on a large scale may be lower than that of the Galactic disc (Morris 1989), and that young OB stars are not the dominant source of infrared luminosity (Cox & Laureijs 1989). In order to reveal the sources of luminosity and heating mechanisms, it is essential to study the energy budget of the clouds on a large scale.

The far-infrared [C II] fine structure line ($^2P_{3/2} \to {}^2P_{1/2}$, 157.7409 $\mu$m, Cooksy, Blake, & Saykally 1986) is a particularly useful probe to study the energy budget of neutral clouds, since the line is the dominant coolant of neutral interstellar gas (Tielens & Hollenbach 1985), and one of the brightest emission lines in the Galaxy (Shibai et al. 1991; Wright et al. 1991). Previous observations of the [C II] line toward the Galactic center, however, were limited to small regions (e.g., Genzel et al. 1985; Poglitsch et al. 1991; Mizutani et al. 1994), or had very low angular resolution (Wright et al. 1991; Bennett et al. 1994). We have made a large-scale survey of the [C II] 158 $\mu$m line emission from the Galactic plane using a balloon-borne telescope (Nakagawa et al. 1993). Our observation covers a wide area (-10°< $l$ < 25°, $|b| \leq 3°$) with sufficiently good spatial resolution (15′) to reveal the basic structure of the Galactic center clouds. Details of our two-dimensional map will be discussed in a forthcoming paper (Nakagawa et al. 1995). In this letter, we concentrate on the longitudinal distribution of the [C II] line emission, particularly the deficit in the ratio of the [C II] line emission to the far-infrared continuum emission at the Galactic center. Origins of this low ratio and their implications are briefly discussed.

## 2. Observation

We observed the far-infrared [C II] line emission from the Galactic plane with the Balloon-borne Infrared Carbon Explorer (BICE, Nakagawa 1993). An offset, oversized optical configuration is adopted for the BICE telescope to reduce the instrumental background radiation, which generally limits the sensitivity of far-infrared observations. The BICE focal plane instrument is a tandem Fabry-Perot spectrometer cooled by liquid helium to 2 K incorporated with a stressed Ge:Ga photoconductor (Hiromoto et al. 1989). The velocity resolution ($\Delta v$) is 175 km s$^{-1}$, and the beam size is 12′.4 (FWHM) with an effective solid angle of $1.5 \times 10^{-5}$ ster. The system NEP during the observation was $6 \times 10^{-16}$ W Hz$^{-1/2}$.

The balloon was launched on 1991 June 12, at the National Scientific Balloon Facility in Palestine, Texas, USA. The floating altitude was 37 - 38 km and the float duration was about 8 hours. The observation time of the Galactic plane was about 6 hours.

In order to observe extended [C II] line emission efficiently as well as to cancel the background radiation, we used a fast spectral scanning method (Yui et al. 1993; Mochizuki et al. 1994) instead of the conventional spatial chopping method. The observation covers roughly $-10° \lesssim l \lesssim 25°$ and $|b| \lesssim 3° \sim 4°$. Instrumental and atmospheric emission was removed by assuming that the astronomical emission is negligible at $|b| > 3°$. The final map was spatially smoothed to $15'$ (FWHM).

We observed M17 as a [C II] flux calibrator during the same flight. The [C II] emission at the peak of M17 within our beam was estimated to be $1.4 \times 10^{-8}$ ergs s$^{-1}$ cm$^{-2}$ on the basis of the M17 [C II] map by Matsuhara et al. (1989). The calibration uncertainty is 30 %, and the detection limit of the final smoothed map is $1.5 \times 10^{-5}$ ergs s$^{-1}$ cm$^{-2}$ ster$^{-1}$ ($3\sigma$).

## 3. Results

In this letter, we concentrate on the longitudinal distributions (Fig.1), which were obtained by averaging the signals over $|b| \leq 2°$ at each longitude. Fig.1a shows our observed longitudinal distribution of the far-infrared [C II] line emission ($I_{\rm [CII]}$). Also shown are the longitudinal distribution of far-infrared continuum emission ($I_{\rm FIR}$, Fig.1b) and that of CO (1-0) line emission (Dame et al. 1987) ($I_{\rm CO}$, Fig.1c).

The [C II] distribution is quite different from others. Although the Galactic center is the dominant peak both in the far-infrared continuum (Fig.1b) and in the CO line (Fig.1c), it is not in the [C II] line (Fig.1a). This situation is better illustrated in Fig.1d, which shows the $I_{\rm [CII]}/I_{\rm FIR}$ ratio; the ratio is almost constant (0.6 %) along the Galactic disk but is systematically low (down to 0.2 %) around the Galactic center. The longitudinal range of the dip is very wide, covering roughly $-3° \lesssim l \lesssim 5°$ (-450 pc to +750 pc).

## 4. Discussion

### 4.1. Optical Depth of the [C II] Line

In this subsection, we estimate the optical depth of the [C II] line toward the Galactic center for the following discussion. Assuming a high density ($n_H \gg 10^3$ cm$^{-3}$) and high temperature ($T \gg 91$ K) limit, the [C II] optical depth is given by (Crawford et al. 1985)

$$\tau_{[CII]} \sim 0.80 \left(\frac{N_H}{10^{21} \text{ cm}^{-2}}\right) \left(\frac{\delta_{C+}}{3 \times 10^{-4}}\right) \left(\frac{1 \text{ km s}^{-1}}{\Delta v}\right) \left(\frac{91 \text{ K}}{T}\right), \qquad (1)$$

where $\delta_{C+}$ is the abundance of C$^+$ ions relative to hydrogen. We take the total hydrogen column density toward the Galactic center as $N_{H,total} = 5 \times 10^{22}$ cm$^{-2}$ (assuming $A_V = 27$ mag, Oort 1977; $N_H/A_V = 1.9 \times 10^{21}$ cm$^{-2}$ mag$^{-1}$, Mathis 1990), and assume that half of it is in the Galactic center region and the rest in the Galactic disk. We also take $\delta_{C+} = 3 \times 10^{-4}$ (see Tielens & Hollenbach 1985, and references therein), $\Delta v = 50$ km s$^{-1}$ (Lugten 1986), and $T = 350$ K (Genzel et al. 1985). This yields $\tau_{[CII]} \sim 0.1$. Although there are some uncertainties in this estimate, we conclude that the [C II] line is optically thin in the Galactic center.

On the other hand, the [C II] line in the *Galactic disk* could be optically thick itself. Heiles (1994) estimated that $\tau_{[CII]}$ is greater than unity in at least some directions for at least some velocity range. But, since the velocity width of the gas in the Galactic disk is much narrower than that in the Galactic center, even if a part of the [C II] line emission from the center could be extinguished at some velocity range by the [C II] opacity in the disk, the bulk of the line emission could not be. Hence we conclude that the optical depth effect alone is not sufficient to explain the low $I_{[CII]}/I_{FIR}$ ratio, and that the intrinsic intensity of the [C II] line is weak in the Galactic center.

## 4.2. Low Abundance of $C^+$ Ions

It is also possible that the [C II] line emission is weak (even with a large amount of $C^+$ ions) or that the abundance of $C^+$ ions themselves is low in the Galactic center. Photodissociation regions or photon-dominated regions (PDRs) are neutral (H I or $H_2$) regions whose chemistry and physics are dominated by far-UV (FUV) photons (6 - 13.6 eV). The PDRs have been thought to be a strong source of [C II] emission (e.g., Tielens & Hollenbach 1985; Shibai et al. 1991). The PDRs are characterized mainly by two physical parameters: (1) the incident FUV flux density of $G_0$, which is normalized to the solar neighborhood value ($1.6 \times 10^{-3}$ ergs cm$^{-2}$ s$^{-1}$, Habing 1968), and (2) the number density ($n$) of gas.

Model calculations (Hollenbach, Tielens, & Takahashi 1991) showed that the [C II] line is the dominant coolant in the PDRs with small $n$ and/or with small $G_0$. In high-density regions with large $G_0$, on the other hand, the [C II] line intensity is rather saturated due to its lower excitation energy and lower critical density, and other lines such as [O I] 63 $\mu$m line become the dominant coolants. Moreover, dust grains become positively charged in the regions with large $G_0$, and the total gas heating efficiency, which is mainly due to the dust photoelectric heating, also decreases (Tielens & Hollenbach 1985). Hence the $I_{\rm [CII]}/I_{\rm FIR}$ ratio decreases at the high $G_0$ end. Fig.2 shows the observed correlation between $I_{\rm [CII]}/I_{\rm FIR}$ and the IRAS 60 $\mu$m and 100 $\mu$m flux ratio, $I_{60}/I_{100}$; the latter is a good measure of $G_0$. The $I_{\rm [CII]}/I_{\rm FIR}$ data for the Galactic disk (crosses in Fig.2), especially the upper envelope of the data, is a clear decreasing function of $I_{60}/I_{100}$, as expected from the above discussion.

However, the situation is quite different for the Galactic center data (circles in Fig.2); they clearly do not correlate with the $I_{60}/I_{100}$ ratio. Furthermore, the $I_{\rm [CII]}/I_{\rm FIR}$ ratios are systematically lower than those of the Galactic plane data irrespective of local $G_0$. Hence we conclude that, in the Galactic center, it is not the strong FUV radiation that reduces the $I_{\rm [CII]}/I_{\rm FIR}$ ratio. Rather, *the abundance of $C^+$ ions must be low on a large scale.* In the following, we discuss possible mechanisms that decrease the abundance of $C^+$ ions.

### 4.3. Soft Interstellar Radiation Field

As discussed in §1, there are some lines of evidence which indicate that the current star-forming activity in the Galactic center may be relatively lower than that of the Galactic disk, and that young OB stars may not be the dominant energy source in the center. Cox & Laureijs (1989) estimated the Infrared Excess (IRE, the ratio of infrared luminosity to the Lyman $\alpha$ luminosity) of the Galactic center region ($2° \times 3°$) on the basis of IRAS observations. Most compact sources associated with the center showed IRE $\sim$ 10, which is typical also for disk H II regions. On the other hand, the diffuse component in the center, which dominates the infrared luminosity, shows a much higher IRE $\sim$ 30. Hence they concluded that the dominant heating source for the dust on a large scale is not young OB stars but rather the population of cool stars - K and M giants - which comprises the Galactic nucleus. Previous balloon-borne observations of diffuse far-infrared thermal emission from the Galactic center also suggest a deficiency of O-stars compared to the solar vicinity (Boissé et al. 1981; Odenwald & Fazio 1984).

On the basis of these IRE differences between the center and the disk H II regions, we assume that (1) most of the infrared luminosity of the *disk* is attributed to young OB stars, and that (2) only one third of the infrared luminosity of the diffuse component in the *center* is due to OB stars while two thirds is attributed to cool stars. Radiation from cool stars can heat the dust, but it contains few carbon-ionizing photons. Hence this relative softness of the interstellar radiation field in the Galactic center reduces the $C^+$ abundance and can roughly explain the small $I_{\rm [CII]}/I_{\rm FIR}$ ratio, which is also one third of the Galactic disk value.

### 4.4. Molecular Self-Shielding

The soft interstellar radiation field has an additional effect on the abundance of the $C^+$ ions. FUV photons more energetic than the ionization potential of hydrogen (13.6 eV) are absorbed within H II regions and less energetic FUV photons go into PDRs. Since the ionization potential of carbon (11.3 eV) is lower than 13.6 eV, carbon is easily ionized at the surface of PDRs, where incident FUV radiation is strong. Within the PDRs, the FUV photons are gradually attenuated, and the $C^+$ becomes C or CO at some depth from the surface.

The depth of the $C^+$/C/CO transition zone is determined mainly by two mechanisms: dust extinction and molecular self-shielding (Tielens & Hollenbach 1985). If there are many FUV photons available relative to gas particles ($G_0/n > 10^{-2}$, Hollenbach, Tielens, & Takahashi 1991), the dust extinction determines the depth ($A_V = 2 \sim 4$ mag) of the transition zone. In this case, $C^+$ regions and far-infrared continuum emitting regions have roughly the same thickness. On the other hand, if there are fewer photons available ($G_0/n < 10^{-2}$, Hollenbach, Tielens, & Takahashi 1991), the molecular self-shielding becomes important. The molecular gas density in the Galactic center is constantly high on a large scale (e.g., Bally et al. 1987; Tsuboi et al. 1989), typically $10^4$ cm$^{-3}$, which is one to two orders of magnitudes larger than those of the disk clouds (Güsten 1989). This high density makes molecular self-shielding more effective in the center than in the disk. The softness of the radiation field in the Galactic center region also makes the molecular self-shielding more effective, since fewer photons are available to photodissociate CO and to ionize C (Spaans et al. 1994). The molecular self-shielding moves the $C^+$/C/CO transition layer closer to the surface of the cloud. Then, the $C^+$ regions become thinner, and the $I_{\rm [CII]}/I_{\rm FIR}$ ratio also decreases.

Hence we conclude that, in the Galactic center region, the combination of pervasively high gas density and soft radiation field makes molecular self-shielding more important in determining the position of the $C^+/C/CO$ transition zone than in disk clouds and the molecular self-shielding also decreases the $I_{[CII]}/I_{FIR}$ ratio.

### 4.5. Warm Molecular Gas in the Galactic Center

Although the Galactic center is not the prominent peak in the [C II] line emission, it shows the prominent peak in the CO (1-0) line (Dame et al. 1987), high-$J$ CO line ($J$ = 2-1, 3-2, 4-3, and 5-4; Bennett et al. 1994), and [C I] line (609 $\mu$m and 370 $\mu$m; Bennett et al. 1994) emission. These results indicate that there is a large amount of warm neutral gas in the Galactic center. Excitation studies of symmetric top molecules (e.g., $NH_3$, $CH_3CCH$, and $CH_3CN$) also suggest uniformly high gas temperatures (50 - 100 K) throughout the central 500 pc irrespective of the local environment (Güsten et al. 1985).

Spaans et al. (1994) showed that, due to molecular self-shielding, the gas temperature ($T_{gas}$) at the $C^+/C/CO$ transition zone becomes higher as the effective temperature of the radiation field ($T_{eff}$) decreases. For example, with $n = 10^3$ cm$^{-3}$ and $G_0$ (Spaans et al. defined it in the 2 - 13.6 eV range) = $10^3$, $T_{gas} \sim 10$ K for $T_{eff}$ = 30,000 K, but $T_{gas} \sim 32$ K for $T_{eff}$ = 6,000 K. Hence the soft radiation field and the molecular self-shielding increase the amount of warm molecular gas. The exact nature of the gas-heating mechanism at the Galactic center has been controversial (Güsten 1989; Morris 1989). We suggest here that the combination of the soft-radiation field and the molecular self-shielding in the Galactic center can effectively heat molecular gas on a large scale. This interpretation is consistent with the observed low $I_{[CII]}/I_{FIR}$ ratio in the Galactic center as we report in this letter.

– 11 –

## 5. Conclusions

(1) The longitudinal distribution of the $I_{\rm [CII]}/I_{\rm FIR}$ ratio is systematically low around the Galactic center. The most reasonable explanation of this is that the abundance of the $C^+$ ions themselves is low in the Galactic center on a large scale.

(2) A likely explanation for the low abundance of $C^+$ ions is the soft radiation field at the Galactic center. The $I_{\rm [CII]}/I_{\rm FIR}$ ratio there is reduced as the ratio of C-ionizing photons to the total interstellar photons decreases.

(3) An additional possible mechanism is molecular self-shielding, which protects a large amount of CO molecules from photodissociation. Molecular self-shielding becomes important in the Galactic center partly due to pervasively high molecular gas number density and partly due to the soft radiation field.

(4) We suggest that the combination of the soft radiation field and the effective molecular self-shielding can produce a large amount of warm molecular gas in the Galactic center and therefore is possibly the effective heating mechanism of the molecular gas.

We are indebted to the staff of the National Scientific Balloon Facility, Palestine, Texas (flight number 1501p). The flight became possible through the efforts of L. J. Caroff and M. D. Bicay at NASA headquarters. We are also grateful to N. Hiromoto for providing us with far-infrared detectors, to N. Yajima and M. Narita for their efforts in developing the BICE system, and to H. Murakami for inspiring discussions. This work was supported by grants-in-aid from Ministry of Education, Science, and Culture in Japan, and by NASA .

Fig. 1.— The longitudinal distributions of (a) $I_{\rm [CII]}$ (this work), (b) $I_{\rm FIR}$ (IRAS), (c) $I_{\rm CO}$ (Dame et al. 1987), and (d) $I_{\rm [CII]}/I_{\rm FIR}$. The distributions are calculated by averaging signals at $|b| \leq 2°$. $I_{\rm FIR}$ is the total flux between 40 - 120 $\mu$m derived from the IRAS 60 $\mu$m and 100 $\mu$m data (Helou et al. 1988). The spatial resolution is $15'$ for (a), (b) and (d), and is $30'$ for (c).

Fig. 2.— The correlation between $I_{\rm [CII]}/I_{\rm FIR}$ and $I_{60}/I_{100}$. Crosses are for the Galactic disk, and circles are for the Galactic center. This plot is made from the longitudinal distribution of $I_{60}$, $I_{100}$, and $I_{\rm [CII]}$ calculated as in Fig.1. The Galactic center region is defined as $-3° \leq l \leq 5°$.